\documentclass{stacs_proc} 
\usepackage{amsfonts} 
\usepackage{graphicx} 
\usepackage{enumerate}
\usepackage{path}
\usepackage{subfig}
\usepackage{hyperref} 

\newenvironment{pf}{\proof}{\qed}
\newcommand{\Tceil}[1]{\ensuremath{\lceil#1\rceil}}
\newcommand{\Tfloor}[1]{\ensuremath{\lfloor#1\rfloor}}

\newcommand{\N}{\ensuremath{\mathbb{N}}}
\newcommand{\R}{\ensuremath{\mathbb{R}}}
\newcommand{\TbbbR}{\ensuremath{\mathbb{R}}}

\newcommand{\eps}{\ensuremath{\varepsilon}}

\def\Gb{\overline{G}}
\def\Tsl{L}
\def\Tsub#1{_{\mbox{\scriptsize\rm #1}}}

\begin{document}
\title[Trimming of Graphs]{Trimming of Graphs, \\[0.8ex] 
  with Application to Point Labeling}

\author[inst1]{T.~Erlebach}{Thomas Erlebach}

\address[inst1]{Department of Computer Science, University of
  Leicester, Leicester LE1 7RH, England.}  
\email{t.erlebach@mcs.le.ac.uk}

\author[inst2]{T.~Hagerup}{Torben Hagerup}

\address[inst2]{Institut f\"ur Informatik, Universit\"at Augsburg,
  86135 Augsburg, Germany.}
\email{hagerup@informatik.uni-augsburg.de}

\author[inst3]{K.~Jansen}{Klaus Jansen}

\address[inst3]{Institut f\"ur Informatik und Praktische Mathematik,
  Universit\"at Kiel, 24098 Kiel, Germany.}
\email{kj@informatik.uni-kiel.de}

\author[inst4]{M.~Minzlaff}{Moritz Minzlaff}

\address[inst4]{Institut f\"ur Mathematik, Technische Universit\"at
  Berlin, 10623~Berlin, Germany.}
\email{minzlaff@math.tu-berlin.de}

\author[inst5]{A.~Wolff}{Alexander Wolff}

\address[inst5]{Faculteit Wiskunde en Informatica, Technische
  Universiteit Eindhoven, the Netherlands.}
\urladdr{www.win.tue.nl/\~{}awolff}



\thanks{Work supported by grant WO~758/4-2 of the German Research
  Foundation (DFG)}

\begin{abstract}
  For $t,g>0$, a vertex-weighted graph of total weight $W$ is
  \emph{$(t,g)$-trimmable} if it contains a
  vertex-induced subgraph
  of total weight at least $(1-1/t)W$ and with
  no simple path of more than $g$ edges.
  A family of graphs is \emph{trimmable} if for each
  constant $t>0$, there is a constant $g=g(t)$ such that
  every vertex-weighted graph in the family is $(t,g)$-trimmable.
  We show that every family of graphs of bounded
  domino treewidth is trimmable.
  This implies that every family of graphs of bounded degree
  is trimmable if the graphs in the family have
  bounded treewidth or are planar.
  Based on this result, we
  derive a polynomial-time
approximation scheme for the problem of labeling 
  weighted points with nonoverlapping
  sliding labels of unit height and given lengths
  so as to maximize the total weight of
  the labeled points.
  This settles one of the last major open questions in the
  theory of map labeling.
\end{abstract}

\maketitle

\stacsheading{2008}{265-276}{Bordeaux}
\firstpageno{265}

\section{Introduction}

\subsection{Graph Trimming}
In this paper we investigate the problem 
of deleting vertices from a given graph so as to
ensure that all simple paths in the remaining
graph are short.
We assume that each vertex has a nonnegative
weight, and we want to delete vertices of
small total weight.
Whereas there is an extensive literature on
separators, which can be viewed as serving to
destroy all large connected components,
we are not aware of previous work on vertex sets
that destroy all long simple paths.
Let us make our notions precise.

\begin{definition}
  For $t>0$ and $g\ge 0$,
  a \emph{$(t,g)$-trimming} of a
  vertex-weighted graph $G=(V,E)$ of total weight $W$ is a set
  $U\subseteq V$ of weight at most $W/t$ such that every simple path
  in $G$ of more than $g$ edges contains a vertex in~$U$.  If $G$
  has a $(t,g)$-trimming, we also say that $G$ is
  \emph{$(t,g)$-trimmable}.
\end{definition}

We say that a family of graphs is \emph{trimmable} if, for every
constant $t>0$, there is a constant~$g\ge 0$
(that depends only on $t$)
such that every vertex-weighted graph in the family is
$(t,g)$-trimmable.
Of course, it suffices to demonstrate this for $t$ larger
than an arbitrary constant.
Not every family of graphs is
trimmable.  For example, if $n,t\ge 2$ and we delete a
$(1/t)$-fraction of the vertices in an unweighted $n$-clique $K_n$, the
remaining graph still has a simple path of $n(1-1/t)-1$ edges.
This expression is not bounded by a function of $t$ alone,
so the family of complete graphs is not trimmable.

With a little effort, one can show the family of
trees to be trimmable.
One popular generalization of trees is based
on the definition below.
Given a graph $G=(V,E)$ and a set $U\subseteq V$,
we denote by $G[U]$ the subgraph of $G$ induced by~$U$.
The \emph{union} of graphs $G_i=(V_i,E_i)$,
for $i=1,\ldots,m$, is the graph
$\bigcup_{i=1}^m G_i=(\bigcup_{i=1}^m V_i,\bigcup_{i=1}^m E_i)$.

\begin{definition}
  A \emph{tree decomposition} of an undirected graph $G=(V,E)$ is a
  pair $(T,B)$, where $T=(X,E_T)$ is a tree and $B:X\to 2^V$ maps each
  node $x$ of $T$ to a subset of $V$, called the \emph{bag} of $x$,
  such that
  \begin{itemize}
  \item $\bigcup_{x\in X}G[B(x)]=G$, and
  \item for all $x,y,z\in X$, if $y$ is on the path from $x$ to $z$ in
    $T$, then $B(x)\cap B(z)\subseteq B(y)$.
  \end{itemize}
  The \emph{width} of the tree decomposition $(T,B)$ is $\max_{x\in
    X}|B(x)|-1$, and the \emph{treewidth} of $G$ is the smallest width
  of any tree decomposition of~$G$.
\end{definition}

This standard definition is given, e.g., by
Bodlaender~\cite{b-pkagb-98}.
The family of graphs of treewidth at most~1
coincides with the family of forests.
By analogy with several other generalizations from the
family of trees to families of graphs of bounded treewidth,
it seems natural to ask whether every family of graphs
of bounded treewidth is trimmable.
At present we cannot answer this question;
we need a concept stronger than bounded treewidth alone.

\begin{definition}
  The \emph{elongation} of a tree decomposition $(T,B)$ is the maximum
  number of edges on a simple path in $T$
  between two nodes with intersecting
  bags.  For every $s\ge 0$,
  let the \emph{$s$-elongation treewidth}
  of an undirected graph $G$ be the smallest width of a tree
  decomposition of $G$ with elongation at most $s$.
\end{definition}

Since every graph has a trivial tree decomposition of
elongation~0, the $s$-elongation treewidth of every graph is
well-defined for every $s\ge 0$.
The 1-elongation treewidth is the \emph{domino
treewidth} studied, e.g., by Bodlaender~\cite{b-ndt-99}.

Our main result about graph trimming, proved in
Section~\ref{sec:trim}, is that for all fixed
$s\ge 0$, every family of graphs
of bounded $s$-elongation treewidth is trimmable.
Ding and Oporowski~\cite{DinO95} proved that
the domino treewidth of a graph can be bounded by a function of its
usual treewidth and its maximum degree.  It follows that every family
of graphs of bounded treewidth and bounded degree is also trimmable.
We derive from this that all families of planar graphs of bounded
degree are trimmable as well.
This result has applications
described below.

\subsection{Label Placement}
Our main motivation for investigating trimmable
graph families arose in the context of
labeling maps with sliding labels.
Generally speaking, map labeling is the problem
of placing a set of labels,
each in the vicinity of the object that
it labels, while meeting certain
conditions.  For an overview, see the map-labeling bibliography
\cite{ws-mlb-96}.  First of all, labels are not allowed to overlap.
As a consequence, it may not be possible to label
all objects in a map, and the goal is to make an
optimal selection according to some criterion.
When a point feature such as a town or a mountain top is to be
labeled, the label can usually be approximated without much loss by an
axes-parallel
rectangular shape and must be placed in the plane without rotation so
that its boundary touches the point.  One distinguishes between
\emph{fixed-position models} and \emph{slider models}.  In
fixed-position models, each label has a predetermined finite set of
\emph{anchor points} on its boundary (e.g., the four corner points),
and the label must be placed so that one of its anchor points
coincides with the point to be labeled.  In slider models, the anchor
points form \emph{anchor segments}
on the boundary of the label (e.g., its bottom
edge).

Van Kreveld et al.~\cite{ksw-plsl-99} introduced a taxonomy of
fixed-position and slider models, which was later refined by Poon et
al.~\cite{pssuw-lpw-03}.
We use the slider models 1SH, 2SH and 4S of Poon et al.,
which define the anchor segments of a label to be its
bottom edge, its top and bottom edges,
and its entire boundary, respectively.
We always require labels to be unit-height rectangles.
This models the case in which
all labels contain single text lines of the same character height.
Fig.~1 illustrates the 1SH model.
We assume that each point to be labeled
comes equipped with a nonnegative weight, which
may be used
to express priorities among the points.  If points represent villages,
towns and cities on a map, priorities may correspond to the number of
inhabitants, for example.  Our objective is to label points with
nonoverlapping labels so as to maximize the sum of the weights of
those points that actually receive a label.  This objective function
causes points with large weights (e.g., large cities) to be likely to
be labeled.
We refer to the specific map-labeling problems described in this
paragraph as
\emph{weighted unit-height 1SH-labeling}, etc.
Since the qualifiers ``weighted'' and ``unit-height''
apply throughout the paper, we may occasionally omit them.

\begin{figure}[t]
  \centering
  \includegraphics{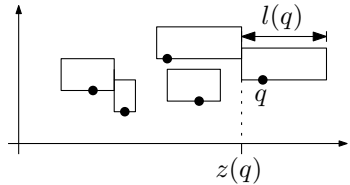}
  \caption{A 1SH-labeling $\Tsl$}
  \label{fig:labeling}
\end{figure}

Recall that for $\rho\le 1$, a \emph{$\rho$-approximation algorithm}
for a maximization problem is an algorithm that
always outputs a solution of value at least $\rho$ times the optimal
objective value.  An algorithm that takes an additional parameter
$\eps>0$ and, for each fixed $\eps$, is a
polynomial-time $(1-\eps)$-approximation
algorithm is called a \emph{polynomial-time
approximation scheme} (\emph{PTAS}).
If the running time depends polynomially on
$\eps$ as well, the algorithm is a \emph{fully polynomial-time
approximation scheme} (\emph{FPTAS}).

Poon et al.~\cite{pssuw-lpw-03} show that finding an optimal weighted
unit-height 1SH-labeling is NP-hard, even if all points lie on a
horizontal line and the weight of each point equals the length of its
label.  For the one-dimensional case, in which all points lie on a
horizontal line, they give an FPTAS,
which yields an $O(n^2/\eps)$-time $(1/2-\eps)$-approximation
algorithm for the two-dimensional case for
arbitrary $\eps>0$.
Poon et al.\ also describe a
PTAS for unit-square labels.  They raise the question of whether a
PTAS exists for rectangular labels of arbitrary length and unit
height.  This is known to be the case for fixed-position models
\cite{aks-lpmis-98} and for sliding labels of unit weight
\cite{ksw-plsl-99}.  The corresponding $(1-\eps)$-approximation
algorithms run in $n^{O(1/\eps)}$ and $n^{O(1/\eps^2)}$ time,
respectively, for arbitrary $\eps>0$.
The question of whether the combination of both sliding
labels and arbitrary weights allows a PTAS has been one of the last
major open problems in (theoretical point-feature) map labeling.
In a preliminary version of this paper \cite{ehjmw-naalw-06}, we made
some progress in answering this question.  We gave a $(2/3 -
\eps)$-approximation for weighted unit-height 1SH-labeling with running time
$n^{O(1/\eps^2)}$, for arbitrary $\eps>0$,
and showed that the same approach yields a PTAS if
the ratio of longest to shortest label length is bounded.

In Section~\ref{sec:labeling} we
settle the open question of Poon et al.\ by
presenting a PTAS for weighted unit-height 1SH-labeling.  There are no
restrictions on
label weights and lengths.
Our approach is to discretize a given instance~$I$ of the
weighted unit-height 1SH-labeling problem, i.e.,
to turn it into a fixed-position instance $I'$,
after which we can apply a known
fixed-position algorithm to~$I'$.
The main difficulty is to find a ``suitable'' set of discrete label
positions for each point.  ``Suitable'' means that
the weight of an optimal labeling of $I'$ must be close enough to
the weight of an optimal labeling of $I$.
Dependencies between labels can be modeled via a graph,
and long simple paths in this graph translate into large sets of
anchor points that cannot be left out of consideration.
Here our results from Section~\ref{sec:trim}
come into play.
We prove that the family of dependency graphs,
if carefully defined, is trimmable, and we show how this
may be used to bound the number of anchor points
by a polynomial.
We also show how to extend our PTAS for
(weighted unit-weight) 1SH-labeling to
the related 2SH-labeling and 4S-labeling problems.

\section{Trimming of Graphs}
\label{sec:trim}%

In this section we show that for every constant~$s$, every family of
graphs of bounded $s$-elongation treewidth is trimmable.
This implies that every family of graphs of
bounded degree is trimmable if the graphs in the family
have bounded treewidth or are planar.

\begin{theorem}
  \label{thm:dom}%
Let $k,s\ge 0$ and
suppose that a vertex-weighted undirected graph $G$
has a tree decomposition $D$ of width $k$ and
elongation $s$.
Take $a=k+1$ if $s\ge 2$ and
$a=\Tceil{k/2}$ if $s\le 1$.
Then, for every integer $t\ge 2$, $G$ has a
$(t,g)$-trimming, where $g=(2(s+1)t-3)(k+1)$ if $a\le 1$ and
\[
g={{(a^{(s+1)t-2}(a+1)-2)(k+1)}/{(a-1)}}
\]
if $a\ge 2$.
Therefore, for every
constant $s$, every family of graphs of bounded
$s$-elongation treewidth is trimmable.
\end{theorem}

\begin{pf}
Let $D=(T,B)$,
root $T$ at an arbitrary node and let $U$ be the set
of vertices in bags whose depth $d$ in $T$ satisfies
$d\bmod(s+1)t=i$, with the integer $i$ chosen
to minimize the weight of $U$.
We show that $U$ is a $(t,g)$-trimming of $G$.

Let $G=(V,E)$ and denote the total weight
of the vertices in $V$ by $W$.
Since each vertex in $V$ occurs in bags on at most
$s+1$ levels in $T$, the sum, over all levels, of the
weight of the vertices occurring in bags on the level
under consideration is at most $(s+1)W$.
Therefore, by the choice of $i$, the
  weight of $U$ is at most ${{(s+1) W}/{((s+1) t)}}={W/t}$,
as desired.

  Let $\pi=(v_0,\ldots,v_m)$ be a simple path in $G$ of
$m\ge 1$ edges and, for
  $i=1,\ldots,m$, choose a node $x_i$ in $T$ whose bag contains
  both $v_{i-1}$ and $v_i$.
Because $T$ is connected, there is a path from $x_i$
to $x_{i+1}$ (or they coincide), for
  $i=1,\ldots,m-1$, so $\pi$ can be viewed as
inducing a walk $\pi'$ in $T$.
The walk $\pi'$ may visit a node $x$ in $T$ several times.
However, each visit to $x$ ``uses'' a vertex in
$B(x)$ that cannot be reused later, so
no node of $T$ occurs more than $k+1$ times on $\pi'$.
If $s\le 1$, we can strengthen this statement as follows:
For $i=1,\ldots,m-1$, the nodes $x_i$ and
$x_{i+1}$ must coincide or be adjacent, so each
visit by $\pi'$ to a node $x$ ``uses'' two
vertices in $B(x)$, rather than just one, and
the number of such
visits is bounded by
$\Tfloor{{(k+1)}/2}=\Tceil{k/2}$.
In either case, therefore, the
  nodes on $\pi'$ span a subtree $T'$ of $T$ in
which no node has more
  than $a$ children, except that the root may have $a+1$
  children.
The number of nodes at depth $d$ in such a tree is
bounded by $(a+1)a^{d-1}$, for all $d\ge 0$, and therefore the
  number of nodes at depth at most $d$ is bounded by $2 d+1$ if $a=1$
  and by $1+{{(a+1)(a^d-1)}/{(a-1)}}=((a+1)a^{d}-2)/(a-1)$ if $a\ge 2$.

  Suppose that $\pi$ contains no vertex in $U$.  Then, by the choice
  of $U$, the depth of~$T'$ is at most $(s+1) t-2$,
and the number of nodes
  in $T'$ is at most $2 (s+1) t-3$ if $a=1$ and at most
$(a^{(s+1)t-2}(a+1)-2)/(a-1)$ if $a\ge 2$.
Since each bag contains at most
  $k+1$ vertices, it follows that $m+1\le(2 (s+1) t-3)(k+1)$ if $a=1$ and
  that $m+1\le {{(a^{(s+1) t-2}(a+1)-2)(k+1)}/{(a-1)}}$ if $a\ge 2$.
\end{pf}

\begin{corollary}
  \label{cor:trim-twdeg}%
  For all integers $k\ge 0$, $d\ge 1$ and $t\ge 2$,
every vertex-weighted undirected graph of
  treewidth $k$ with maximum degree $d$ has a $(t,\Tceil{K/2}^{2
    t})$-trimming, where $K=(9 k+7)d(d+1)-1$.
  Hence, every family of graphs with bounded degree and bounded
  treewidth is trimmable.
\end{corollary}

\begin{pf}
  According to Bodlaender \cite[Theorem~3.1]{b-ndt-99}, 
  every such graph has a domino tree
  decomposition of width at most~$K$.
Except in the trivial case $k=0$, we have $K\ge 31$.
  By Theorem~\ref{thm:dom}, used with
$s=1$, the graph has a
  $(t,g)$-trimming, where
  \[
  g={{(\Tceil{K/2}^{2 t-2}(\Tceil{K/2}+1)-2)(K+1)}/{(\Tceil{K/2}-1)}}
  \le\Tceil{K/2}^{2 t}.
  \]
\end{pf}

We can extend this result to planar graphs of bounded degree.

\begin{corollary}
  \label{cor:trim-planar}%
  For all integers $d,t\ge 1$,
  every vertex-weighted undirected planar graph of
  maximum degree $d$ has a $(t,\Tceil{K/2}^{4 t})$-trimming, 
  where $K=(54t-29)d(d+1)-1$.
  Hence every family of planar graphs of
bounded degree is trimmable.
\end{corollary}

\begin{pf}
  Let $G=(V,E)$ be a planar graph with
maximum degree $d$
  and denote the total weight of the vertices in $V$ by~$W$.
  We first follow the approach of Baker~\cite{b-aancp-94} to
  obtain a $(2t-1)$-outerplanar subgraph of~$G$ by deleting
  vertices of total weight at most~$W/(2t)$.
  Consider an arbitrary planar embedding of $G$.
  Partition the vertices of $G$ into layers by repeatedly
  deleting the vertices on the
  boundary of the outer face
  until no vertex remains.
  The vertices deleted
  in one iteration of this process form a layer.
  Number the layers $R_1,R_2,\ldots$ in the order of
  their deletion. For every
  $j\in\{0,1,\ldots,2t-1\}$, consider the
  set $V_j$ of vertices in layers $R_i$
with $i\bmod(2t)=j$, choose $j$
such that the total weight
  of $V_j$ is at most~$W/(2t)$
and consider the subgraph $H_j$ of $G$
  induced by $V\setminus V_j$.

  $H_j$ is $(2t-1)$-outerplanar and
  thus has treewidth at most~$6t-4$~\cite[Theorem 83]{b-pkagb-98}.
  By Corollary~\ref{cor:trim-twdeg},
  $H_j$ has a $(2t,\Tceil{K/2}^{4 t})$-trimming~$U$.
  The set $V_j\cup U$ has weight at most
  $W/(2t) + W/(2t) = W/t$ and is therefore a
  $(t,\Tceil{K/2}^{4 t})$-trimming of~$G$.
\end{pf}

\begin{remark}
A better dependence of the
bound in Corollary~\ref{cor:trim-planar} on~$t$
can be achieved by deleting less than $1/(2t)$
of the weight of the graph in the first step,
so that more than $1/(2t)$ of the weight can be
deleted when Corollary~\ref{cor:trim-twdeg}
is applied. In this way, the treewidth
of $H_j$ and thus the value of $K$ increases,
but the exponent of the bound becomes
smaller than~$4t$. More precisely, if we delete $1/(\alpha t)$ of the
weight in the first step, for some $\alpha>2$, then the resulting
bound is $\Tceil{K/2}^{2\Tceil{\alpha t/(\alpha-1)}}$ with
$K = (27\alpha t-29)d(d+1)-1$.
For each pair $(d,t)$,
there is a value of $\alpha$ that optimizes the resulting
bound.
\end{remark}

\section{Labeling Weighted Points with Sliding Labels}
\label{sec:labeling}%

In this section we define the labeling
problems of relevance to us formally and show
that there are polynomial-time
approximation schemes for weighted
unit-height 1SH-labeling, 2SH-labeling and 4S-labeling.
We use $\TbbbR$, $\TbbbR_{>0}$ and $\TbbbR_{\ge 0}$
to denote the sets of real numbers, of positive
real numbers and of nonnegative real numbers,
respectively, and
$\TbbbR^2$ is the two-dimensional Euclidean plane.

\begin{definition}
An instance of the
\emph{weighted unit-height 1SH-labeling problem}
is a triple $I=(P,l,w)$, where $P$ is a finite
subset of $\TbbbR^2$ and $l:P\to\TbbbR_{>0}$
and $w:P\to\TbbbR_{\ge 0}$ are functions
defined on~$P$.
$|P|$ is called the \emph{size} of~$I$.
\end{definition}

In the definition of 1SH-labeling, $P$ represents the
set of points to be labeled, and for each $p\in P$,
$l(p)$ is the length of the label of $p$
and $w(p)$ is the weight of $p$.
When $(P,l,w)$ is an instance of the
1SH-labeling problem
and $Q\subseteq P$, we call
$w(Q)=\sum_{p\in Q}w(p)$ the \emph{weight} of~$Q$.

\begin{definition}
\label{def:labeling}
A feasible solution or \emph{labeling} of
an instance $I=(P,l,w)$ of the
weighted unit-height 1SH-labeling problem
is a pair $\Tsl=(Q,z)$, where $Q\subseteq P$
and $z:Q\to\R$ is a function with
$p_x-l(p)\le z(p)\le p_x$ for all $p=(p_x,p_y)\in Q$
such that for all $p=(p_x,p_y)$
and $q=(q_x,q_y)$ in $Q$ with $p\not=q$
and $|p_y-q_y|<1$, either
$z(p)+l(p)\le z(q)$ or $z(q)+l(q)\le z(p)$.
The \emph{weight} of $\Tsl$ is the weight of~$Q$,
and $\Tsl$ is \emph{optimal} if
no labeling of $I$ has greater weight than $\Tsl$.
\end{definition}

Informally, $Q$ is the set of points in $P$
that receive a label, and for each $p\in Q$,
$z(p)$ denotes the $x$-coordinate of the left
edge of the label of~$p$.
The condition $p_x-l(p)\le z(p)\le p_x$ for all
$p=(p_x,p_y)\in Q$ expresses that $p$ lies on
the bottom edge of its label.
Let us say that two points $p=(p_x,p_y)$
and $q=(q_x,q_y)$ in $\TbbbR^2$
\emph{$y$-overlap} if $|p_y-q_y|<1$.
The condition $z(p)+l(p)\le z(q)$ or $z(q)+l(q)\le z(p)$
for each pair $(p,q)$ of distinct $y$-overlapping
points in $Q$ expresses that labels are not allowed to overlap.

We define an instance of the
\emph{weighted unit-height
multi-position labeling}
or \emph{1MH-labeling problem}
as a pair $(I,\mathcal{M})$, where $I=(P,l,w)$ is an
instance of the
weighted unit-height 1SH-labeling problem
and $\mathcal{M}$ is a function that maps each
point in $P$ to a finite subset of $\TbbbR$.
A \emph{labeling} of $(I,\mathcal{M})$ is a labeling
$(Q,z)$ of $I$ such that $z(p)\in \mathcal{M}(p)$
for all $p\in Q$.
If~$\mathcal{M}$ maps all $p\in P$ to the same
set $M\subseteq\TbbbR$, we may write
$(I,\mathcal{M})$ as $(I,M)$.
The principal technical contribution of this
section is a reduction of 1SH-labeling to 1MH-labeling.
Before giving a precise description of the
reduction, we provide an informal overview.

The reduction maps an instance $I$ of 1SH-labeling
to an instance of 1MH-labeling of the form $(I,M)$,
where $M\subseteq\TbbbR$.
It therefore suffices to show that a suitable
set $M$ exists and can be computed
sufficiently fast.
As a step towards this goal,
we describe a \emph{normalization} procedure
that transforms an arbitrary given labeling of $I$
into one of $(I,M)$.
The normalization is introduced for the
sake of argument only and is not actually carried out
as part of the reduction.

The top-level idea behind the
normalization is to process
the labels of the given labeling in the order from left
to right, pushing each label
as far to the left as it can go without
bumping into another label or
being separated from the point that it labels.
It is easy to observe that in every normalized labeling,
the position of each label (taken to be the
$x$-coordinate of its left edge) is the sum of
the $x$-coordinate of some labeled point and some
number of label lengths, minus its own length.
This still leaves too many possibilities, however,
since essentially every selection of points to
receive labels may give rise to a different
position of a given label.

The dependencies between labels can be modeled in
a natural way through a directed
\emph{dependency graph} $G$:
If the label of a point $q$, moving left, may bump into
that of a point~$p$, then~$G$ includes the edge $(p,q)$.
The problem identified above stems from the
fact that~$G$ may have very long paths,
corresponding to chains of many
labels that may touch and influence each other.
Our defense against this is trimming, so we
must ensure that $G$ is trimmable.
Assuming that this is so, we can break all paths
with more than a constant number of edges
by dropping labels of small total weight,
which reduces the number of possible label
positions to a polynomial.
Afterwards we must re-normalize, however, since
otherwise the trimming buys us nothing.
This gives rise to another problem, in that
the re-normalization may create new long paths.
In order to counter this, we introduce vertical
\emph{stopping lines} and modify the normalization
to never push the left edge of a label past a stopping line.
As long as at least one stopping line passes through
each dropped label (including its boundary), we can be sure
that the re-normalization creates no new paths.
Fairly arbitrarily, for every label,
we choose to put stopping lines
through the left and right edges of the area occupied
by the label in its leftmost position
(if no other labels obstruct its movement).
This also ensures in a simple way that no label
gets separated from the point that it labels.
Now labels with their right edge to the left of
or on a stopping line $\ell$ cannot influence labels
with their left edge to the right of or on~$\ell$,
so we can remove all edges from $G$ 
that cross a stopping line.
This turns out to have the beneficial effect of
making $G$ planar and of bounded degree, which
implies that it is trimmable, as needed above.

By attaching real-valued lengths to the
edges of $G$ and adding an additional 
vertex~$O$ with incident edges described below to $G$,
we can obtain the position of the label of each point~$p$
as the length of a path from $O$ to~$p$.
Every edge $(p,q)$ between two points $p$ and $q$
is given a length equal to that of the label of $p$,
since that is the distance that the left edge of
the label of~$q$
must keep from that of~$p$.
Every stopping line $\ell$, passing through $(x,0)$, say,
and every point $p$ give rise to an edge from
$O$ (which can be thought of as representing
the $y$-axis) to $p$ of length $x$,
since $x$ is the distance that the left
edge of the label of $p$, because of
$\ell$, must keep from the $y$-axis if it begins
its movement to the right of $\ell$ or on $\ell$.
Now the label of each point
$p$ will move to a position that is precisely
the largest length of a path from~$O$ to~$p$
no larger than the original position of the label.

Every stopping line adds to the number of possible
label positions in a normalized labeling,
but the dependence on the number of stopping lines
is only linear.
In fact, because of a later need for
this added flexibility, Lemma~\ref{lem:labeling} below
allows the specification of an
arbitrary set $S$ of $x$-coordinates of additional
stopping lines.
The fact that the left edge of a label
crosses no additional stopping line as it moves left
can be expressed by saying that the movement
leaves the rank in $S$ of the position of the label
invariant.

\begin{lemma}
\label{lem:labeling}
Given an instance $I=(P,w,l)$ of the
weighted unit-height 1SH-labeling
problem of size $n$,
a finite set $S\subseteq\TbbbR$
and an $\eps\in\TbbbR$ with $0<\eps\le 1$,
in $O((n+|S|)n^g)$ time,
where $g=({1/\eps})^{O({1/\eps})}$,
we can compute a set
$M\subseteq\TbbbR$ with $|M|\le (2 n+|S|)n^g$
such that for every labeling $(Q,z)$ of $I$,
the instance $(I,M)$ of
the weighted unit-height
1MH-labeling problem
has a labeling $(Q',z')$ with $Q'\subseteq Q$
of weight at least $(1-\eps)w(Q)$
such that for all $p\in Q'$,
$z'(p)\le z(p)$ and $z'(p)$ and $z(p)$
have the same rank in~$S$.
\end{lemma}

\begin{pf}
Take $S'=S\cup\bigcup_{(p_x,p_y)\in P}\{p_x-l(p),p_x\}$
and let $G=(Q,E)$ be the directed
graph with edge lengths
on the vertex set $Q$ that,
for all $p=(p_x,p_y)$ and $q=(q_x,q_y)$ in
$Q$, contains the edge $(p,q)$ with length $l(p)$ exactly if
$p_x<q_x$, $|p_y-q_y|<1$ and there is
no $x\in S'$ with $z(p)+l(p)\le x\le z(q)$.
Moreover, let $H$ be the undirected graph
on the vertex set $Q$ that contains an edge
$\{p,q\}$, for all $p,q\in Q$ with $p\not=q$,
exactly if $p$ and $q$ $y$-overlap.

Let us say that two points $p=(p_x,p_y)$ and $r=(r_x,r_y)$
in $Q$ \emph{$x$-surround}
a point $q=(q_x,q_y)$ if
$p_x\le q_x\le r_x$ or $r_x\le q_x\le p_x$.
Let $p$, $q=(q_x,q_y)$ and $r$ be three points in $Q$, every
two of which $y$-overlap, and suppose that
$z(p)\le z(q)\le z(r)$.
Then we must clearly have
$z(p)+l(p)\le z(q)\le q_x\le z(q)+l(q)\le z(r)$,
which, since $q_x\in S'$, implies that $(p,r)\not\in E$.
This proves the following \emph{triangle property}:
If $(p,q)\in E$, then $p$ and $q$ $x$-surround
no neighbor of both in~$H$.

If $p=(p_x,p_y)\in Q$,
then all in- and out-neighbors of $p$ in $G$
lie in the open horizontal strip of height 2
centered on the line $y=y_p$.
Therefore, if $p$ has in- or out-degree 3 or more,
two in-neighbors or two out-neighbors of $p$
are neighbors in $H$, which contradicts
the triangle property.
Thus all in- and out-degrees of $G$
are bounded by~2.

We next prove that $G$ is planar.
Consider an embedding
of $G$ that maps each point in~$Q$ to itself and
each edge in $E$ to a straight line segment and
assume to the contrary that for two edges
$(p_1,q_1)$ and $(p_2,q_2)$ in $E$ with
$|\{p_1,q_1,p_2,q_2\}|=4$, the corresponding closed
line segments $\overline{p_1 q_1}$ and $\overline{p_2 q_2}$
intersect in a point $u=(u_x,u_y)$.
Call $p_1$ and $q_1$ as well as $p_2$ and $q_2$
\emph{partners} and let $H_4$ be the subgraph of $H$
spanned by the vertex set $Q_4=\{p_1,q_1,p_2,q_2\}$.

All points in $Q_4$ lie in the open
horizontal strip
of height 2 centered on the line $\ell$
defined by $y=u_y$.
If there are a topmost point in $Q_4$
(one of maximal $y$-coordinate)
and a bottommost point in $Q_4$ that are
partners, then, since these $y$-overlap,
all pairs of points in $Q_4$ $y$-overlap,
and $H_4$ is a complete graph.
Otherwise there is a unique topmost point
and a unique bottommost point in $Q_4$,
these \emph{extreme} points
are not partners,
and each of the two other points
in $Q_4$ lies on $\ell$
or on the opposite side of $\ell$ with respect
to its extreme partner.
Each nonextreme point in $Q_4$ $y$-overlaps
both extreme points, and hence
also the fourth point in $Q_4$,
either by virtue of lying on $\ell$
or because one extreme point is its
partner, while the other extreme point
lies on the same side of $\ell$ as itself.
This means that
$H_4$ is a complete graph, except that the
two extreme points may not be neighbors.

Because the two line segments between partners
intersect, some two points in $Q_4$ that are partners,
say, $a$ and $b$,
must $x$-surround another point in $Q_4$,
say, $c$.
By the triangle property, $H$ lacks one of the
edges $\{a,c\}$ and $\{b,c\}$, say, $\{b,c\}$,
so $H$ is not complete and $b$ and $c$ are extreme.
The partner of $c$, say, $d$, is not extreme,
so it is not $x$-surrounded by $a$ and $b$.
This implies that $c$ and $d$ $x$-surround
$a$ or $b$ and, in fact, since $a$ is
not extreme, that they $x$-surround $b$.
The two extreme points $b$ and $c$
can now be seen
to be $x$-surrounded by $a$ and $d$.
But then it is geometrically clear that $a$
and $d$ belong to opposite open halfspaces
bounded by the line through $b$ and $c$
(see Fig.~\ref{fig:planar}),
a contradiction to the fact
that $\overline{a b}$ and $\overline{c d}$ intersect.

\begin{figure}
  \centering
  \includegraphics{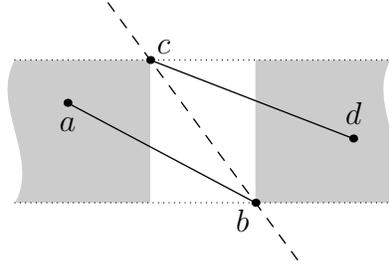}
  \caption{$a$ and $d$ lie in distinct gray
 areas and are therefore on opposite sides of $\overline{b c}$.}
  \label{fig:planar}
\end{figure}

We have demonstrated that $G$ is planar and
of bounded degree and therefore trimmable.
With $t=2/\eps$,
let $U$ be a $(t,g)$-trimming set of $G$ for some
integer $g\ge 0$ with $g=t^{O(t)}$---this is possible
by Corollary~\ref{cor:trim-planar}---and
take $Q'=Q\setminus U$.
Let $\Gb$ be the multigraph obtained from~$G$
by adding a new vertex $O$ and, for each $x\in S'$
and each $p\in Q$,
an edge from~$O$ to~$p$ of length~$x$.

For all $p\in Q'$, let a \emph{$p$-path}
be a path in $\Gb[\{O\}\cup Q']$ from $O$ to~$p$
and define the \emph{length} of a $p$-path
as the sum of the lengths of its edges.
For all $p=(p_x,p_y)\in Q'$, let $z'(p)$ be
the largest length of a $p$-path
that does not exceed $z(p)$---this is well-defined since
$z(p)\ge p_x-l(p)$, while there is an edge, and
hence a path, in $\Gb$ from $O$ to $p$ of length $p_x-l(p)$.
We will show that $(Q',z')$ is a
labeling of $I$.
First, for each $p=(p_x,p_y)\in Q'$,
the relation $p_x-l(p)\le z'(p)\le z(p)\le p_x$
was essentially argued above.
Second, we must show, informally speaking,
that the labels of the points
in $Q'$, if placed as indicated by $z'$, do not overlap.

Let $p=(p_x,p_y)$ and $q=(q_x,q_y)$ be
$y$-overlapping points in $Q'$
and assume, without loss of generality,
that $z(p)\le z(q)$ and therefore that $z(p)+l(p)\le z(q)$.
If $G$ contains the edge $(p,q)$, then,
since $z'(p)$ is the length of a $p$-path,
$z'(p)+l(p)$ is the length of a $q$-path and,
by definition of $z'$, we have $z'(q)\ge z'(p)+l(p)$.
If $G$ does not contain the edge $(p,q)$,
there is an $x\in S'$ with $z(p)+l(p)\le x\le z(q)$.
Again by definition of $z'$, since $\Gb$ contains
an edge from $O$ to $q$ of length $x$, it follows that
$z'(q)\ge x\ge z(p)+l(p)\ge z'(p)+l(p)$.
In either case, the labels of $p$ and $q$,
placed according to $z'$, do not overlap.

We have $w(Q')\ge(1-{1/t})w(Q)$, and for each $p\in Q'$,
$z'(p)$ is the length of a $p$-path.
The length of every $p$-path belongs to
the set $M$ of all sums of an element of $S'$
and at most~$g$ elements of $\{l(p)\mid p\in P\}$.
The set $M$ is of size at most $(2 n+|S|)n^g$
and can be computed in $O((n+|S|)n^g)$ time.
Let $p\in Q'$.
Since for each $x\in S$ there is a $p$-path
of length $x$, it is easy to see that
stepping from $z(p)$ to $z'(p)$ does not
descend strictly below any $x\in S$, i.e.,
$z'(p)$ has the same rank in $S$ as $z(p)$.
\end{pf}

We need to show how to solve the instance of
the 1MH-labeling problem obtained using
Lemma~\ref{lem:labeling}.
Agarwal et al.~\cite{aks-lpmis-98} have given a PTAS that finds
near-maximum independent sets in any given set of axes-aligned
unit-height rectangles. They assume that rectangles are topologically
closed. Under this assumption it is easy to argue that their PTAS for
maximum independent set at the same time is a PTAS for maximizing the
number of points labeled with unit-height
rectangular labels in some fixed-position model.  The reason is simply
that, by definition, any two label candidates of the same point must 
touch this point. If label candidates are closed, one label candidate
automatically excludes the other from the solution. Unfortunately,
this is not the case if we consider labels to be open; e.g., in the 1SH-model the
leftmost and the rightmost label candidate of a point do \emph{not}
intersect, so an algorithm for maximum independent set would not
automatically yield feasible solutions for multi-position labeling.
However, we can adapt the PTAS of Agarwal et
al.\ to this case. In fact, the adapted PTAS
can deal with the \emph{weighted unit-height generalized multi-position
labeling} or \emph{4M-labeling problem}, in which each label specifies
an arbitrary finite set of anchor points on its boundary.
If a point is labeled, its label must be placed so that one of
its anchor points coincides with the point to be labeled.

\begin{lemma}
  \label{lem:open}
  There is a PTAS for the weighted unit-height 4M-labeling problem.
  The running time for computing a $(1-\eps)$-approximate solution
  is $n^{O(1/\eps)}$, for all $\eps$ with $0<\eps\le 1$.
\end{lemma}

Clearly, a PTAS for 4M-labeling is also a PTAS for the more
restricted 1MH-labeling problem.

\begin{theorem}
Given an instance $I$ of the
weighted unit-height 1SH-labeling
problem of size $n$
and an $\eps\in\TbbbR$ with $0<\eps\le 1$,
a labeling of $I$ of
weight at least $(1-\eps)$ times the
weight of an optimal labeling of~$I$ can be computed
in $n^{t^{O(t)}}$ time, where $t={2/\eps}$.
The weighted unit-height 1SH-labeling problem
therefore admits a PTAS.
\end{theorem}

\begin{pf}
Let $W^*$ be the weight of an optimal
labeling of~$I$.
Use the algorithm of Lemma~\ref{lem:labeling} with $S=\emptyset$
to compute a set $M\subseteq\TbbbR$
with $|M|\le 2 n^{g+1}$,
where $g=t^{O(t)}$,
such that the instance
$I'=(I,M)$ of
the weighted unit-height 1MH-labeling problem
has a labeling of weight at least $(1-{1/t})W^*$.
Applying the PTAS of Lemma~\ref{lem:open}
to $I'$, we obtain a labeling of $I'$,
and therefore of $I$, of weight at least
$(1-{1/t})^2 W^*\ge(1-{2/t})W^*=(1-\eps)W^*$
in time $(n^{g+2})^{O(t)}=n^{t^{O(t)}}$, which
dominates the time needed by the first step.
\end{pf}

This result can be extended without much effort to the slightly more
general labeling model 2SH, where a label must touch the point
labeled with either its top or bottom edge.

\begin{corollary}
  There is a PTAS for weighted unit-height 2SH-labeling.
\end{corollary}

\begin{pf}
  2SH-labeling can be reduced to 1SH-labeling---imagine adding to each
  original input point a copy at a distance of~1 below it. Then we use the
  reduction from 1SH-labeling to 1MH-labeling described in
  Lemma~\ref{lem:labeling}. In the resulting instance of 1MH-labeling,
  we discard the copies of points and view each label of a copy of
  a point as labeling the original point. Now we can apply the
  PTAS of Lemma~\ref{lem:open} to the resulting instance of 4M-labeling.
\end{pf}

A further generalization allows us to deal also with the most general
slider model, 4S, in which a label may have the point
that it labels anywhere
on its boundary.

\begin{corollary}
  There is a PTAS for weighted unit-height 4S-labeling.
\end{corollary}

\begin{Proof}[Proof sketch.]
Let an instance $I=(P,l,w)$ of the 4S-labeling problem
(which is the same as an instance of the 1SH-labeling problem)
be given.
Each point $p\in P$ can be labeled with a horizontally sliding label
that touches $p$ with its bottom edge (or top edge), or by a vertically
sliding label that touches $p$ with its left edge (or right edge).
This means that there are four types of rectangles that can potentially
label~$p$, all of which are taken into account in the following.
Applying Lemma~\ref{lem:labeling} twice (once horizontally and
once vertically),
we compute an instance $I\Tsub h$ of the 1MH-labeling problem
for the positions of horizontally
sliding labels, specifying vertical stopping lines at $x$-positions
$p_x-l(p)$, $p_x$ and $p_x+l(p)$ for all $p=(p_x,p_y)$ in $P$,
and another instance $I\Tsub v$ for the positions of vertically sliding
labels, specifying horizontal stopping lines at $y$-positions
$p_y-1$, $p_y$ and $p_y+1$ for all $p=(p_x,p_y)$ in~$P$.
Consider an optimal labeling $L$ of $I$
and let $Q$ be the set of points that it
labels. Let $Q\Tsub h$ and $Q\Tsub v$ be the sets of points
in $Q$ that are labeled with a horizontally sliding label
and with a vertically sliding label, respectively.
By Lemma~\ref{lem:labeling}, there
is a solution~$L\Tsub h'$ for~$I\Tsub h$
that labels points $Q\Tsub h'\subseteq Q\Tsub h$,
and a solution~$L\Tsub v'$ for~$I\Tsub v$
that labels points $Q\Tsub v'\subseteq Q\Tsub v$,
of weights at least $(1-\eps)w(Q\Tsub h)$ and $(1-\eps)w(Q\Tsub v)$,
respectively. Furthermore, the labels in $Q\Tsub h'$ reach their positions
in $L\Tsub h'$ from their position in $L$ by sliding horizontally without crossing
a vertical stopping line. Thus, they do not interfere with the vertical
movement that vertically sliding labels undergo in the transition from
$L$ to $L\Tsub v'$, and vice versa. Consequently, the union of~$L\Tsub h'$
and~$L\Tsub v'$ (defined in the obvious way)
is a labeling of $I$ of weight
at least $(1-\eps)$ times the optimum. Applying 
the PTAS of Lemma~\ref{lem:open}
to $I\Tsub h\cup I\Tsub v$, we obtain a solution
of~$I$ of weight at least
$(1-\eps)w(Q\Tsub h'\cup Q\Tsub v')\ge (1-\eps)^2 w(Q)$,
which completes the proof.
\end{Proof}

\section{Open Problems}
Corollary~\ref{cor:trim-twdeg} states that a family of graphs
is trimmable if it is of bounded treewidth
\emph{and} bounded degree.
We cannot exclude, however, that the bounded-degree
condition is superfluous.
In other words, with $\N=\{1,2,\ldots\}$, is
there a function $g:\N\times\N\to\N$ such that for
  all $k,t\in\N$, \emph{every} weighted undirected graph of treewidth~$k$
  has a $(t,g(k,t))$-trimming?
The answer is yes in the unweighted case, i.e., if
all weights are the same.
If the answer were generally yes,
it would follow by the argument
in the proof of Corollary~\ref{cor:trim-planar} that the family of
planar graphs is also trimmable.
More generally,
the question of which families of graphs are trimmable
deserves further study.

\section*{Acknowledgments}

We thank Hans Bodlaender for pointing us to the concept
of domino treewidth.


\newcommand{\etalchar}[1]{$^{#1}$}


\begin{thebibliography}{vKSW99}

\bibitem[AvKS98]{aks-lpmis-98}
Pankaj~K. Agarwal, Marc van Kreveld, and Subhash Suri.
\newblock Label placement by maximum independent set in rectangles.
\newblock {\em Comput. Geom. Theory Appl.}, 11:209--218, 1998.

\bibitem[Bak94]{b-aancp-94}
Brenda~S. Baker.
\newblock Approximation algorithms for {NP}-complete problems on planar graphs.
\newblock {\em J. ACM}, 41:153--180, 1994.

\bibitem[Bod98]{b-pkagb-98}
Hans~L. Bodlaender.
\newblock A partial $k$-arboretum of graphs with bounded treewidth.
\newblock {\em Theoret. Comput. Sci.}, 209(1--2):1--45, 1998.

\bibitem[Bod99]{b-ndt-99}
Hans~L. Bodlaender.
\newblock A note on domino treewidth.
\newblock {\em Discrete Math. Theor. Comput. Sci.}, 3(4):141--150, 1999.

\bibitem[DO95]{DinO95}
Guoli Ding and Bogdan Oporowski.
\newblock Some results on tree decomposition of graphs.
\newblock {\em J.\ Graph Theory}, 20:481--499, 1995.

\bibitem[EHJ{\etalchar{+}}06]{ehjmw-naalw-06}
Thomas Erlebach, Torben Hagerup, Klaus Jansen, Moritz Minzlaff, and Alexander
  Wolff.
\newblock A new approximation algorithm for labeling weighted points with
  sliding labels.
\newblock In {\em Proc. 22nd European Workshop on Computational Geometry
  (EWCG'06)}, pages 137--140, Delphi, 2006.

\bibitem[PSS{\etalchar{+}}03]{pssuw-lpw-03}
Sheung-Hung Poon, Chan-Su Shin, Tycho Strijk, Takeaki Uno, and Alexander Wolff.
\newblock Labeling points with weights.
\newblock {\em Algorithmica}, 38(2):341--362, 2003.

\bibitem[vKSW99]{ksw-plsl-99}
Marc van Kreveld, Tycho Strijk, and Alexander Wolff.
\newblock Point labeling with sliding labels.
\newblock {\em Comput. Geom. Theory Appl.}, 13:21--47, 1999.

\bibitem[WS96]{ws-mlb-96}
Alexander Wolff and Tycho Strijk.
\newblock {The Map-Labeling Bibliography}.
\newblock \url{http://i11www.ira.uka.de/map-labeling/bibliography}, 1996.

\end{thebibliography}
\end{document}